\documentclass[preprint,superscriptaddress,showpacs,prl,aps,letterpaper]{revtex4}

\usepackage{natbib}
\usepackage{epsf}
\usepackage{graphicx}

\newcommand{\ppbar}{$p\overline{p}$~}

\def\comment#1{}

\def\withcomments{
\addtolength{\oddsidemargin}{-0.5 in}
\addtolength{\evensidemargin}{-0.5 in}
\newcounter{mycommentcounter}
\def\comment##1{\refstepcounter{mycommentcounter}%
  \ifhmode%
  \unskip%
  {\dimen1=\baselineskip \divide\dimen1 by 2 %
    \raise\dimen1\llap{\tiny -\themycommentcounter-}}\fi%
  \marginpar{\renewcommand{\baselinestretch}{0.8}%
    \footnotesize [\themycommentcounter]: \raggedright ##1}}
\date{\framebox{Draft of \today}}
}

\def\simge{\mathrel{%
   \rlap{\raise 0.511ex \hbox{$>$}}{\lower 0.511ex \hbox{$\sim$}}}}
\def\simle{\mathrel{
   \rlap{\raise 0.511ex \hbox{$<$}}{\lower 0.511ex \hbox{$\sim$}}}}
\def\lessim{\mathrel {\vcenter {\baselineskip 0pt \kern 0pt
\hbox{$<$} \kern 0pt \hbox{$\sim$} }}}
\def\gessim{\mathrel {\vcenter {\baselineskip 0pt \kern 0pt
\hbox{$>$} \kern 0pt \hbox{$\sim$} }}}
\def \rightdownarrow
 {\kern.4em {\raise1.75ex\hbox{$\Bigl |$}$\kern-0.27em{\longrightarrow}$}}
\def \mrightdownarrow
 {\kern.4em {\raise1.25ex \hbox{$|$} \kern-0.27em{
                                                \longrightarrow}}}

\def\mMEt{\not\kern-.35em {E_T}}

\begin{document}

\title{\boldmath{ Search for High-Mass Resonances Decaying to $e\mu$ in $p\bar{p}$ Collisions at $\sqrt{s}=1.96$ TeV }}\unboldmath
\affiliation{Institute of Physics, Academia Sinica, Taipei, Taiwan 11529, Republic of China} 
\affiliation{Argonne National Laboratory, Argonne, Illinois 60439} 
\affiliation{Institut de Fisica d'Altes Energies, Universitat Autonoma de Barcelona, E-08193, Bellaterra (Barcelona), Spain} 
\affiliation{Baylor University, Waco, Texas  76798} 
\affiliation{Istituto Nazionale di Fisica Nucleare, University of Bologna, I-40127 Bologna, Italy} 
\affiliation{Brandeis University, Waltham, Massachusetts 02254} 
\affiliation{University of California, Davis, Davis, California  95616} 
\affiliation{University of California, Los Angeles, Los Angeles, California  90024} 
\affiliation{University of California, San Diego, La Jolla, California  92093} 
\affiliation{University of California, Santa Barbara, Santa Barbara, California 93106} 
\affiliation{Instituto de Fisica de Cantabria, CSIC-University of Cantabria, 39005 Santander, Spain} 
\affiliation{Carnegie Mellon University, Pittsburgh, PA  15213} 
\affiliation{Enrico Fermi Institute, University of Chicago, Chicago, Illinois 60637} 
\affiliation{Joint Institute for Nuclear Research, RU-141980 Dubna, Russia} 
\affiliation{Duke University, Durham, North Carolina  27708} 
\affiliation{Fermi National Accelerator Laboratory, Batavia, Illinois 60510} 
\affiliation{University of Florida, Gainesville, Florida  32611} 
\affiliation{Laboratori Nazionali di Frascati, Istituto Nazionale di Fisica Nucleare, I-00044 Frascati, Italy} 
\affiliation{University of Geneva, CH-1211 Geneva 4, Switzerland} 
\affiliation{Glasgow University, Glasgow G12 8QQ, United Kingdom} 
\affiliation{Harvard University, Cambridge, Massachusetts 02138} 
\affiliation{Division of High Energy Physics, Department of Physics, University of Helsinki and Helsinki Institute of Physics, FIN-00014, Helsinki, Finland} 
\affiliation{University of Illinois, Urbana, Illinois 61801} 
\affiliation{The Johns Hopkins University, Baltimore, Maryland 21218} 
\affiliation{Institut f\"{u}r Experimentelle Kernphysik, Universit\"{a}t Karlsruhe, 76128 Karlsruhe, Germany} 
\affiliation{High Energy Accelerator Research Organization (KEK), Tsukuba, Ibaraki 305, Japan} 
\affiliation{Center for High Energy Physics: Kyungpook National University, Taegu 702-701; Seoul National University, Seoul 151-742; and SungKyunKwan University, Suwon 440-746; Korea} 
\affiliation{Ernest Orlando Lawrence Berkeley National Laboratory, Berkeley, California 94720} 
\affiliation{University of Liverpool, Liverpool L69 7ZE, United Kingdom} 
\affiliation{University College London, London WC1E 6BT, United Kingdom} 
\affiliation{Centro de Investigaciones Energeticas Medioambientales y Tecnologicas, E-28040 Madrid, Spain} 
\affiliation{Massachusetts Institute of Technology, Cambridge, Massachusetts  02139} 
\affiliation{Institute of Particle Physics: McGill University, Montr\'{e}al, Canada H3A~2T8; and University of Toronto, Toronto, Canada M5S~1A7} 
\affiliation{University of Michigan, Ann Arbor, Michigan 48109} 
\affiliation{Michigan State University, East Lansing, Michigan  48824} 
\affiliation{Institution for Theoretical and Experimental Physics, ITEP, Moscow 117259, Russia} 
\affiliation{University of New Mexico, Albuquerque, New Mexico 87131} 
\affiliation{Northwestern University, Evanston, Illinois  60208} 
\affiliation{The Ohio State University, Columbus, Ohio  43210} 
\affiliation{Okayama University, Okayama 700-8530, Japan} 
\affiliation{Osaka City University, Osaka 588, Japan} 
\affiliation{University of Oxford, Oxford OX1 3RH, United Kingdom} 
\affiliation{University of Padova, Istituto Nazionale di Fisica Nucleare, Sezione di Padova-Trento, I-35131 Padova, Italy} 
\affiliation{LPNHE, Universite Pierre et Marie Curie/IN2P3-CNRS, UMR7585, Paris, F-75252 France} 
\affiliation{University of Pennsylvania, Philadelphia, Pennsylvania 19104} 
\affiliation{Istituto Nazionale di Fisica Nucleare Pisa, Universities of Pisa, Siena and Scuola Normale Superiore, I-56127 Pisa, Italy} 
\affiliation{University of Pittsburgh, Pittsburgh, Pennsylvania 15260} 
\affiliation{Purdue University, West Lafayette, Indiana 47907} 
\affiliation{University of Rochester, Rochester, New York 14627} 
\affiliation{The Rockefeller University, New York, New York 10021} 
\affiliation{Istituto Nazionale di Fisica Nucleare, Sezione di Roma 1, University of Rome ``La Sapienza," I-00185 Roma, Italy} 
\affiliation{Rutgers University, Piscataway, New Jersey 08855} 
\affiliation{Texas A\&M University, College Station, Texas 77843} 
\affiliation{Istituto Nazionale di Fisica Nucleare, University of Trieste/\ Udine, Italy} 
\affiliation{University of Tsukuba, Tsukuba, Ibaraki 305, Japan} 
\affiliation{Tufts University, Medford, Massachusetts 02155} 
\affiliation{Waseda University, Tokyo 169, Japan} 
\affiliation{Wayne State University, Detroit, Michigan  48201} 
\affiliation{University of Wisconsin, Madison, Wisconsin 53706} 
\affiliation{Yale University, New Haven, Connecticut 06520} 
\author{A.~Abulencia}
\affiliation{University of Illinois, Urbana, Illinois 61801}
\author{D.~Acosta}
\affiliation{University of Florida, Gainesville, Florida  32611}
\author{J.~Adelman}
\affiliation{Enrico Fermi Institute, University of Chicago, Chicago, Illinois 60637}
\author{T.~Affolder}
\affiliation{University of California, Santa Barbara, Santa Barbara, California 93106}
\author{T.~Akimoto}
\affiliation{University of Tsukuba, Tsukuba, Ibaraki 305, Japan}
\author{M.G.~Albrow}
\affiliation{Fermi National Accelerator Laboratory, Batavia, Illinois 60510}
\author{D.~Ambrose}
\affiliation{Fermi National Accelerator Laboratory, Batavia, Illinois 60510}
\author{S.~Amerio}
\affiliation{University of Padova, Istituto Nazionale di Fisica Nucleare, Sezione di Padova-Trento, I-35131 Padova, Italy}
\author{D.~Amidei}
\affiliation{University of Michigan, Ann Arbor, Michigan 48109}
\author{A.~Anastassov}
\affiliation{Rutgers University, Piscataway, New Jersey 08855}
% ...
\collaboration{CDF Collaboration}
\noaffiliation

\date{\today}

\begin{abstract}
We describe a general search for resonances decaying to a neutral $e\mu$ final state in \ppbar collisions at a center-of-mass energy of 1.96~TeV.  Using a data sample representing 344 pb$^{-1}$ of integrated luminosity recorded by the CDF~II experiment, we compare Standard Model predictions with the number of observed events for invariant masses between $50$ and $800$ GeV/$c^2$.  Finding no significant excess ( $5$ events observed vs. $7.7 \pm 0.8$ expected for $M_{e\mu} > 100$ GeV/$c^2$ ), we set limits on sneutrino and $Z'$ masses as functions of lepton family number violating couplings.
\end{abstract}


\maketitle

An observed excess of high-mass opposite sign electron-muon pairs at the Tevatron would provide evidence of physics beyond the Standard Model (SM).  Nearly every extension of the SM predicts flavor changing neutral current effects.    General supersymmetric models, for instance, permit the violation of R-parity symmetry (RPV) and describe the lepton family number violating (LFV) production and decay of sneutrinos ($\tilde{\nu}$), the scalar superpartners of SM neutrinos~\cite{YanBin}.  Models with additional gauge symmetry can also accommodate an $e\mu$ signature through LFV decays of a new heavy neutral gauge boson, the $Z'$~\cite{murakami}.  Signals from such processes may be easily detected at the Tevatron.  SM backgrounds are small and characterized by invariant mass spectra that lie well below the range presumed for new physics.  

We search for a general signature of high-mass opposite sign $e\mu$ pairs and are sensitive to both resonant and non-resonant production mechanisms.  The CDF collaboration investigated the high-mass $e\mu$ channel in a Run~I search for the direct RPV production and decay of the tau sneutrino ($\tilde{\nu}_{\tau}$)~\cite{tamburello}~\cite{dzero}.  That analysis excluded sneutrino masses below $\sim360$ GeV/$c^2$ for a particular set of RPV coupling values.  This Letter describes our continued investigation of this channel using data from Run~II of the Tevatron taken with the upgraded Collider Detector at Fermilab (CDF~II).  The data sample we use represents an integrated luminosity of 344 $\pm$ 21~pb$^{-1}$, approximately three times that used in Run~I.  We search for an  $e\mu$ resonance by examining invariant masses ($M_{e\mu}$) between $100$ and $800$ GeV/$c^2$ for an excess of events.  We find event counts that are consistent with SM predictions and use our result to constrain models of LFV $\tilde{\nu}$ and $Z'$ decay by excluding $e\mu$ coupling values as functions of the new particle masses.  In contrast to searches that utilize low-energy data, where the effects of new heavy particles may be masked by cancellations of various contributions, the interpretation of our results is robust.

The CDF~II detector is an azimuthally and forward-backward symmetric apparatus designed to study \ppbar~reactions at the Tevatron.  The detector has a charged particle tracking system immersed in a 1.4~T magnetic field aligned coaxially with the \ppbar beams.  A 3.1~m long open-cell drift chamber, the Central Outer Tracker (COT)~\cite{cot}, covers the radial range from 40 to 137~cm and provides coverage for the pseudo-rapidity range $|\eta|\simle 1$~\cite{geomnote}. Segmented electromagnetic and hadronic sampling calorimeters surround the tracking system and measure the energies of interacting particles in the range $|\eta|<3.6$~\cite{em,had,plug}.  A set of drift chambers located outside the central hadron calorimeters and another set behind a 60~cm thick iron shield track muons with $|\eta|\le0.6$.~\cite{muon}.  Additional drift chambers and scintillation counters detect muons in the region $0.6\le|\eta|\le1.0$.  Gas Cerenkov counters located in the $3.7<|\eta|<4.7$ region~\cite{clc} measure the average number of inelastic \ppbar collisions  per bunch crossing and thereby determine the beam luminosity.

We use data collected with high-$P_{T}$ triggers that require central
($|\eta|\simle1.0$) lepton candidates.  We select events that contain
at least one reconstructed electron of $E_{T}>20$ GeV and one
reconstructed muon with $P_{T}>20$ GeV/c.  Energy deposited by the leptons in the central electromagnetic calorimeters must be of an isolated nature and should also satisfy the standard set of ``tight'' CDF lepton identification (ID) criteria~\cite{wz}.  

We select oppositely charged $e\mu$ pairs and ensure that the lepton
tracks come from a common $p\bar{p}$ interaction.  We reject events
containing cosmic-ray muons and photon-conversion electrons by
imposing additional event topology requirements~\cite{wz}.  In order
to maximize acceptance to a broad range of new physics signatures, we
do not impose requirements on missing transverse energy or jets.

SM $e\mu$ backgrounds include $q\bar{q}\rightarrow Z/\gamma^{\star}\rightarrow \tau\tau$ (Drell-Yan), top ($t\bar{t}$) and diboson ($WW$, $WZ$, and $ZZ$) production.  We determine geometric and kinematic acceptances for these processes using the PYTHIA Monte Carlo generator~\cite{pythia} and a GEANT3~\cite{geant} based simulation of the CDF~II detector.  These simulations employ the CTEQ5L~\cite{cteq5l} parton distribution functions (PDF's) to model the momentum distribution of the initial state partons.  We define the acceptance for each background process, $\alpha_{i}$, as the fraction of generated events that satisfy the lepton ID and event topology requirements.  We correct the $\alpha_{i}$ by multiplying with trigger efficiencies ($\epsilon_{trg}$) measured from $W\rightarrow e\nu$ and $Z\rightarrow\mu\mu$ data and factors ($f_{reco}$, $f_{ID}$) that account for differences in lepton reconstruction and ID efficiency between simulation and data~\cite{wz}.  We refer to the combined correction factor, $\epsilon_{trg}\times\!f_{reco}\times\!f_{ID}$, as $\epsilon$.  

The expected contribution of each SM background is given by the product of $\alpha_{i}\times\epsilon$ with the corresponding cross section and the integrated luminosity of our data sample.  We estimate top and diboson background contributions using next-to-leading order (NLO) cross sections: $6.1$ pb for $t\bar{t}$~\cite{top}, $12.1$ pb for $WW$, $3.7$ pb for $WZ$ and $1.4$ pb for $ZZ$~\cite{diboson}.  We use a next-to-next-to-leading order continuum ($M_{ll}>30$ GeV/$c^2$) cross section for the Drell-Yan process, $337.7$ pb, which we calculate by scaling the PYTHIA leading order cross section by the ratio of NNLO to LO predictions obtained from PHOZPR calculations~\cite{drellyan}.  

Jets that are misidentified as leptons account for approximately 5\% of the total background.  We determine the misidentification
probability (``fake rate'')~\cite{wz} by examining separate data
samples collected with various jet $E_{T}$ triggers for the occurrence
of leptons.  The real leptonic content of these samples is assumed to
be negligible.  We obtain fake rates from the ratios of misidentified
leptons and the number of jets in these samples.  We then estimate the
background from this source by applying the measured fake rates,
parameterized as functions of $E_{T}$, to the jets in events in the
high-$P_{T}$ sample that contain a single lepton candidate.

\begin{table}[htbp] 
\begin{center}
\begin{tabular}{c|c|c}
\hline
Channel & $50<M_{e\mu}<100$~GeV/$c^2$  & $M_{e\mu}>100$~GeV/$c^2$\\  
\hline
$Z/\gamma^{\star}\rightarrow\tau\tau$ & 38.8 $\pm$ 2.9     & 0.6 $\pm$ 0.0 \\ 
WW,WZ,ZZ                              &  6.6 $\pm$ 0.5     & 3.5 $\pm$ 0.2 \\ 
$t\bar{t}$                            &  3.6 $\pm$ 0.5     & 3.2 $\pm$ 0.5 \\ 
Fake Lepton                           &  2.9 $\pm$ 1.7     & 0.4 $^{+0.6}_{-0.4}$\\
\hline
Prediction                            & 51.9 $\pm$ 3.4     & 7.7 $\pm$ 0.8 \\
Observation                           & 56                 & 5\\
\hline
\end{tabular}
\caption{Expected and observed event totals for low and high-mass $M_{e\mu}$ regions.  The uncertainties shown are the combination of statistical and systematic errors.  }
\label{table:events}
\end{center}
\end{table}
We present the number of observed and predicted background events for two $M_{e\mu}$ regions in Table~\ref{table:events}.  The dominant uncertainty on the background predictions arises from a 6\% uncertainty in the luminosity measurement~\cite{lum}.  Additional uncertainty contributions include those associated with the SM cross sections, fake rate measurements, lepton ID and reconstruction scale factors, and PDF model.  Overall, we find that uncertainties on our signal acceptance and background expectations have little impact on the limits we set.

\begin{figure}[h]
\begin{center}
\epsfxsize=3.0in
\epsfbox{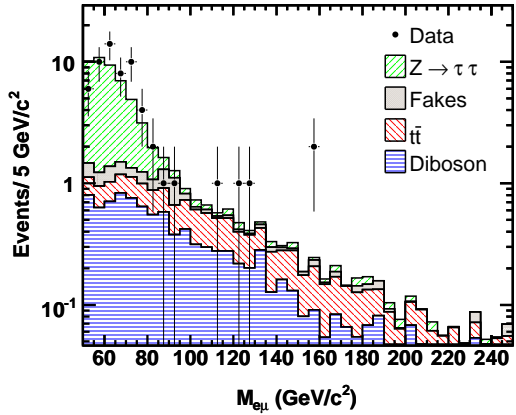}
\caption{Observed and predicted $M_{e\mu}$ Distributions.  The observed $e\mu$ invariant mass spectrum agrees well with that of the combined SM and fake lepton backgrounds.  No events are observed in data beyond $159~\rm{GeV/c^{2}}$.}
\label{fig:mass}
\end{center}
\end{figure}

The $50\leq M_{e\mu}\leq 100$ GeV/$c^2$ region listed in Table~\ref{table:events} represents an invariant mass range that is rich in background.  Finding good agreement between our observation and predicted background in this region, we next consider the $M_{e\mu}$ range above $100$ GeV/$c^2$.  Here, too, we find good agreement.  Figure~\ref{fig:mass} shows the observed and predicted background $M_{e\mu}$ distributions over a portion of the full $50-800$ GeV/$c^2$ invariant mass range.


We quantify the consistency between the $M_{e\mu}$ distributions presented in Fig.~\ref{fig:mass} by performing a $\chi^2$ test, using variable-width $M_{e\mu}$ bins to achieve sufficient predicted background occupancies for the Gaussian approximation of Poisson statistics.  The test results in a total reduced $\chi^2$ statistic ($\chi^2/N_{dof}$) of $14.0/11=1.27$, under the assumption that systematic uncertainties in each bin are completely correlated.  The statistic implies a p-value, the probability of finding a larger $\chi^{2}$, of $\sim23\%$.  This value provides a statistical basis for accepting our results as consistent with SM expectations.

Finding no evidence of new physics in the $M_{e\mu}$ spectrum, we turn from the model-independent search to constraints on specific models; RPV sneutrino and LFV $Z'$ decay.  In RPV supersymmetric models, the s-channel $d\bar{d}\rightarrow\tilde{\nu}_{i}(\bar{\tilde{\nu}}_{i})\rightarrow\!e\mu$ process is governed by two LFV couplings.  $\lambda'_{i11}$ determines the sneutrino production cross section from $d$ and $\bar{d}$ while $\lambda_{1i2}$ gives the sneutrino branching ratio to $e\mu$~\cite{choudhury}.  The index $i$ refers to the lepton generation of the sneutrino.  We do not consider initial-states that include up-type quarks since RPV sneutrino hadro-production occurs only through a $L_{i}Q_{j}\bar{D}_{k}$ term in the superpotential~\cite{choudhury}.  The final-state for this process consists of $e\mu$ only, \emph{i.e.,} it does not contain additional neutrinos or a SUSY LSP.

Since strong limits exist for $\tilde{\nu}_{e}$ and
$\tilde{\nu}_{\mu}$ couplings~\cite{chemtob}, we focus instead on the
third generation sneutrino, $\tilde{\nu}_{\tau}$.  We assume the
``single coupling dominance'' hypothesis~\cite{chemtob} and set all
$\lambda$ and $\lambda'$ couplings but $\lambda'_{311}$ and
$\lambda_{132}$ to zero so that contributions to the $e\mu$ channel
originate from the $\tilde{\nu}_{\tau}$ only.  Previous limits on
$\lambda'_{311}$ and $\lambda_{132}$ from low energy experiments are
$0.10$ and $0.05$~\cite{ledroit}, assuming squark and slepton masses
of $100$ GeV/$c^2$.

\begin{figure}[h]
\begin{center}
\epsfxsize=3.0in
\epsfbox{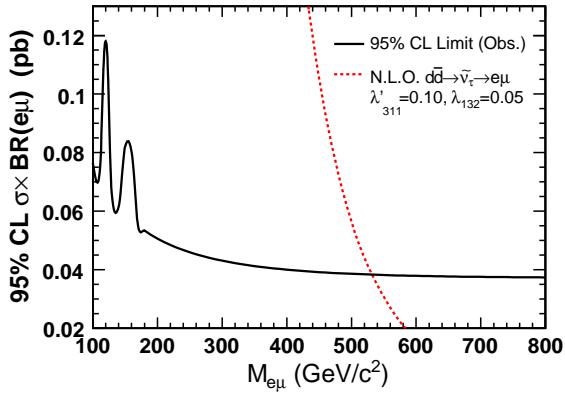}
\caption{Observed 95\% CL upper limit on $\sigma\times\!BR$ for
  $d\bar{d}\rightarrow\tilde{\nu}_{\tau}(\bar{\tilde{\nu}}_{\tau})\rightarrow\!e\mu$
  (solid line) and the NLO prediction (dashed line) as a function of
  $e\mu$ invariant mass.  Their intersection gives a $~530$ GeV/$c^2$
  $\tilde{\nu}_{\tau}$ mass limit for the values of $\lambda'_{311}$
  and $\lambda_{132}$ indicated.  Due to small differences in the
  signal acceptances, the 95\% CL upper limit on the $Z'$
  $\sigma\times\!BR$ (not shown) is larger than that of the sneutrino,
  $\sim 0.02$ pb greater at $200$ GeV/$c^2$ and $\sim 0.003$ pb
  greater at $600$ GeV/$c^2$, for example.}
\label{fig:sigmabr}
\end{center}
\end{figure}

We show an example NLO $\sigma\times\!BR$~\cite{choudhury-personal} that corresponds to $\lambda'_{311}=0.10$ and $\lambda_{132}=0.05$ in Fig.~\ref{fig:sigmabr}.  We use such curves for different $\lambda_{132}$ and $\lambda'_{311}$ values and a mass-dependent $\alpha\times\epsilon$ calculated from PYTHIA Monte Carlo simulations of the $d\bar{d}\rightarrow\tilde{\nu}_{\tau}(\bar{\tilde{\nu}}_{\tau})\rightarrow\!e\mu$ process to obtain signal expectations over the $100$ to $800$ GeV/$c^2$ range.  Assuming coupling values of $\lambda_{132}=0.05$ and $\lambda'_{311}=0.1$, for example, we find that a hypothetical $300$ GeV/$c^2$ $\tilde{\nu}_{\tau}$ signal consists of $16.2\pm1.3$ events.
    
We calculate an upper limit $\sigma\times\!BR$ for $d\bar{d}\rightarrow\tilde{\nu}_{\tau}(\bar{\tilde{\nu}}_{\tau})\rightarrow\!e\mu$ using the numbers of observed and predicted background events in bins of $M_{e\mu}$.  We scan the $M_{e\mu}$ range in steps of 10 GeV/$c^2$ and use the simulated mass resolution at each step to weight all data and background events between $100$ and $800$ GeV/$c^2$.  A Bayesian algorithm~\cite{joel} is used with a uniform prior to translate the weighted event totals to a 95\% confidence level (CL) upper limit on the number of signal events in data.  We divide this limit by the integrated luminosity and signal $\alpha\times\epsilon$ to obtain a 95\% CL upper limit on the
$\sigma\times\!BR$ for the process.  The $\alpha\times\epsilon$ for a
generic spin-0 particle decaying to $e\mu$ is mass-dependent,
increasing slowly from $\sim 10\%$ at $100$ GeV/$c^2$ to $\sim 27\%$
at $800$ GeV/$c^2$.  We account for uncertainties on the signal and
background expectations in limit calculation although the final
results are largely insensitive to these numbers.

\begin{figure}[h]
\begin{center}
\epsfxsize=3.0in
\epsfbox{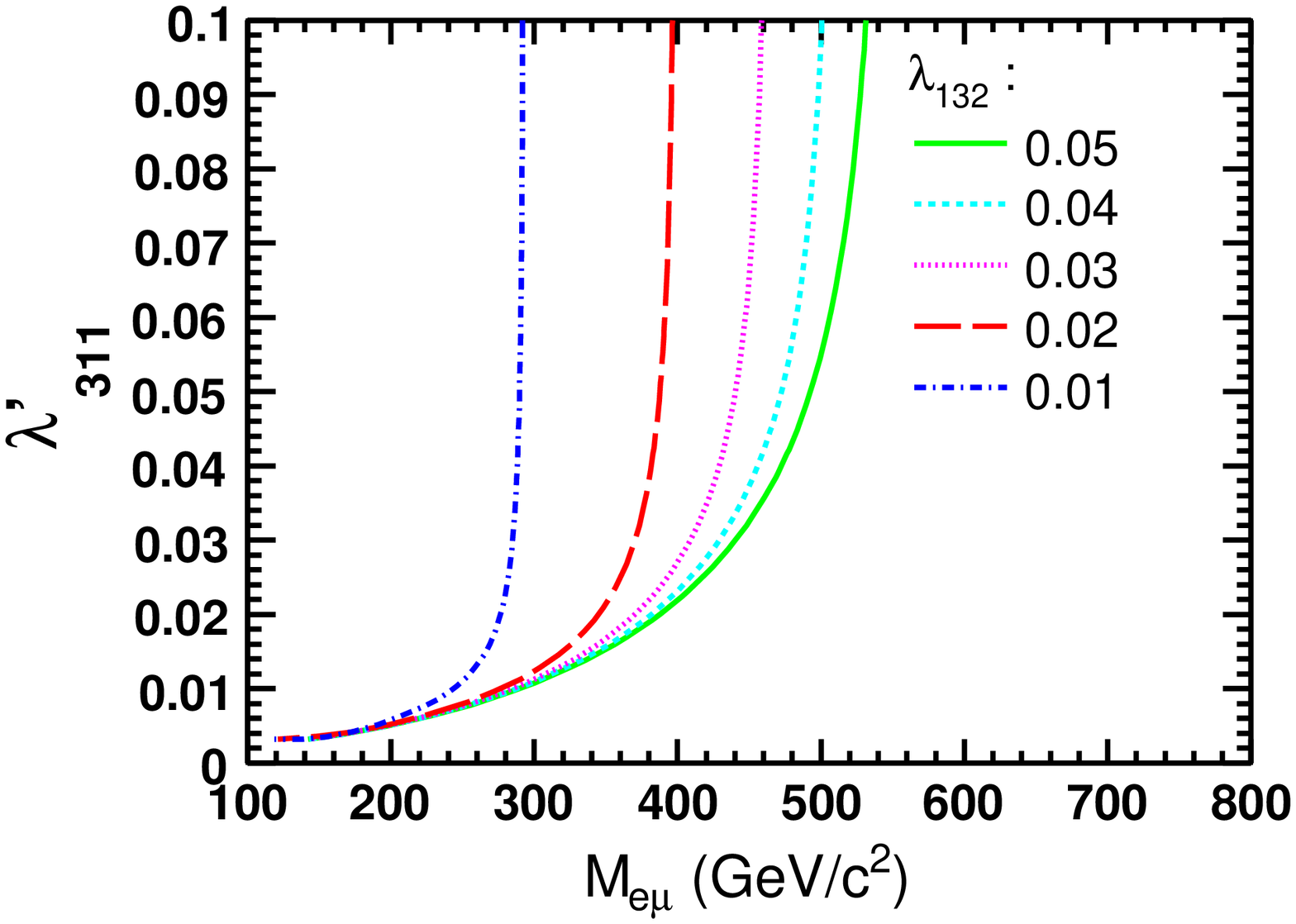}
\caption{ $\lambda'_{311}-M_{e\mu}$ Exclusion Regions.  Regions to the left of the five curves shown represent ranges of $\lambda'_{311}$ values that we exclude at 95\% CL as a function of $\tilde{\nu}_{\tau}$ mass.  Each region corresponds to a fixed value of $\lambda_{132}$.} 
\label{fig:exclusion-snu}
\end{center}
\end{figure}

The intersection of the observed upper limit and a NLO $\sigma\times\!BR$ curves defines a $\tilde{\nu}_{\tau}$ mass limit for specific values of the $\lambda'_{311}$ and $\lambda_{132}$ couplings, as shown in Fig.~\ref{fig:sigmabr}.  We construct exclusion regions in the $\lambda'_{311}$-$M_{e\mu}$ plane by plotting the mass limit as a function of both RPV couplings.  Figure \ref{fig:exclusion-snu} shows the exclusion region parameterized by five assumed values of $\lambda_{132}$.  The plot indicates, for example, that we exclude at 95\% CL $\lambda'_{311}$ values above $\sim 0.01$ for a sample $\tilde{\nu}_{\tau}$ mass of $300$ GeV/$c^2$ and $\lambda_{132}\simge 0.02$.  

Our search is also sensitive to LFV $Z'$ decay.  The
$p\bar{p}\rightarrow\!Z'\rightarrow\!e\mu$ process proceeds through
diagonal $U(1)'$ couplings at the initial-state vertex and an
off-diagonal LFV $U(1)'$ coupling, $Q^{l}_{12}$, that determines the
$Z'$ branching ratio ($BR$) to $e\mu$~\cite{murakami}.  We set 95\% CL
limits on $Q^{l}_{12}$ as a function of the $Z'$ mass using NLO
$\sigma\times\!BR$'s from a group of $E_{6}$-inspired models.
Although $E_{6}$ models do not incorporate the LFV $Q^{l}_{12}$
coupling by construction,
we use these models because they provide a theoretical reference by
specifying initial-state $Z'$-quark coupling and a NLO $Z'$ production
cross section.  We utilize NLO cross sections provided by the $\chi$,
$\psi$, Secluded and $\eta$ $E_{6}$ models~\cite{langacker} and extend
these models to include the $Q^{l}_{12}$
coupling~\cite{langacker-personal}, which introduces
$Z'\rightarrow\!e\mu$ decays.

We set limits on $Q^{l}_{12}$ as a function of the $Z'$ mass following
the procedure previously described for RPV $\tilde{\nu}_{\tau}$ decay.
The $\alpha\times\epsilon$ we use in calculating the upper limit on
$\sigma\times\!BR$ is obtained from PYTHIA Monte Carlo simulations in
which the $Z'$ is required to couple to a left-handed leptonic LFV
current, as may be favored in $E_6$-motivated $U(1)^\prime$ models
that incorporate LFV~\cite{langacker-personal}.  $Z'$ acceptance
($\sim 8\%$ at 100 GeV/$c^2$, and increasing to $\sim 20\%$ at $800$
GeV/$c^2$) is smaller than that for the scalar sneutrino due to the
spin-1 nature of the particle.

\begin{figure}[h]
\begin{center}
\epsfxsize=3.0in
\epsfbox{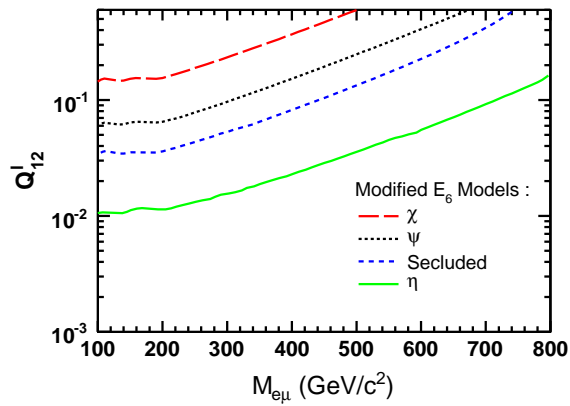}
\caption{ $Q^{l}_{12}-M_{e\mu}$ Exclusion Regions depicting 95\% CL upper limits on the LFV $Q^{l}_{12}$ coupling in extended $\psi$, $\chi$, $\eta$ and Secluded $E_{6}$ models. } 
\label{fig:exclusion-zp}
\end{center}
\end{figure}

The $Q^{l}_{12}$-$M_{e\mu}$ regions that we exclude for the modified $\chi$, $\psi$, Secluded and $\eta$ models are shown in Fig.~\ref{fig:exclusion-zp}.  Assuming initial-state $Z'$ couplings specified by the $\eta$ model, for example, we exclude $Q^{l}_{12}$ values above $\sim0.01-0.1$ for $Z'$ masses between $200$ and $700$ GeV/$c^2$.  Because the LFV $Q^{l}_{12}$ coupling is not intrinsic to $E_{6}$ models, our limits should not be interpreted as constraints on the models themselves.

In summary, we searched in $344$~pb$^{-1}$ of CDF~II data for
high-mass $e\mu$ events and found an $M_{e\mu}$ distribution consistent with SM
predictions.  With no evidence for new physics, we set correlated 95\%
CL limits on the mass and $\lambda'_{311}$ and $\lambda_{132}$ RPV
couplings of the $\tilde{\nu}_{\tau}$.  We achieve a $\sim 170$
GeV/$c^2$ improvement in the $\tilde{\nu}_{\tau}$ mass limit for
specific RPV coupling values of $\lambda'_{311}=0.1$ and
$\lambda_{132}=0.05$~\cite{pdg} and, more generally, exclude
$\tilde{\nu}_{\tau}$ masses over a range of $\lambda'_{311}$ and
$\lambda_{132}$ values.  We also place limits
on potential LFV $Q^{l}_{12}$ couplings of the $Z'$ boson as a
function of its mass using $E_6$-like models of $U(1)'$ symmetry.

We thank the Fermilab staff and the technical staffs of the participating institutions for their vital contributions.  We thank S. Mahji for providing us with NLO $\tilde{\nu}$ production cross sections and B. Dobrescu, J. Kang and P. Langacker for guidance on incorporating LFV $Z'$ decay in $E_{6}$ models.  This work was supported by the U.S. Department of Energy and National Science Foundation; the Italian Istituto Nazionale di Fisica Nucleare; the Ministry of Education, Culture, Sports, Science and Technology of Japan; the Natural Sciences and Engineering Research Council of Canada; the National Science Council of the Republic of China; the Swiss National Science Foundation; the A.P. Sloan Foundation; the Bundesministerium fuer Bildung und Forschung, Germany; the Korean Science and Engineering Foundation and the Korean Research Foundation; the Particle Physics and Astronomy Research Council and the Royal Society, UK; the Russian Foundation for Basic Research; the Comision Interministerial de Ciencia y Tecnologia, Spain; in part by the European Community's Human Potential Programme under contract HPRN-CT-20002, Probe for New Physics; by the Research Fund of Istanbul University Project No. 1755/21122001; and by the Research Corporation. 

\bibliography{emuprl}

\begin{thebibliography}{20}
\bibitem[{geo()}]{geomnote}
\bibinfo{note}{We use a cylindrical coordinate system about the beam axis in
  which $\theta$ is the polar angle, $\phi$ is the azimuthal angle, and $\eta$
  is pseudorapidity, $-\ln \tan (\theta /2)$. $E_T \equiv E \sin \theta$ and $P_T \equiv P
  \sin \theta$ where $E$ is energy measured by the calorimeter and $P$ is
  the magnitude of the momentum measured by the spectrometer. \MET$\equiv -\sum_i E_T^i {\bf n}_i$, where ${\bf n}_i$ is the unit vector in the azimuthal plane that points from the beamline to the $i$th calorimeter tower.}
\end{thebibliography}

\end{document}